\documentclass[a4paper,11pt]{article}
\usepackage{pos}

\usepackage{graphicx} 
\usepackage{physics}
\usepackage{color}
\usepackage{comment}
\usepackage{siunitx}

\title{Calculation of meson charge radii using model-independent method in the PACS10 configuration}

\author*[a]{Kohei Sato}
\author[b]{Hiromasa~Watanabe}
\author[c,d]{Takeshi~Yamazaki}
\affiliation[]{\normalsize{\bf \sffamily \hspace{50mm} (PACS Collaboration)}}

\affiliation[a]{Degree Programs in Pure and Applied Sciences, Graduate School of Science and Technology, University of Tsukuba, Tsukuba, Ibaraki 305-8571, Japan}
\affiliation[b]{Yukawa Institute for Theoretical Physics, Kyoto University, Kyoto 606-8502, Japan}
\affiliation[c]{Faculty of Pure and Applied Sciences, University of Tsukuba, Tsukuba, Ibaraki 305-8571, Japan}
\affiliation[d]{Center for Computational Sciences, University of Tsukuba, Tsukuba, Ibaraki 305-8577, Japan}

\emailAdd{ksatoh@het.ph.tsukuba.ac.jp}

\abstract{
We report our preliminary results for the charge radii of $\pi^{+}$ and $K^{+}$ mesons with the PACS10 configuration generated at the physical point using the Iwasaki gauge action and $N_{f}=2+1$ stout-smeared nonperturbatively $\order{a}$ improved Wilson quark action, especially at $\SI{0.085}{fm}$ corresponding lattice size $128^4$. The charge radii are obtained from a model-independent method that directly calculates the first-order differential coefficient of the electromagnetic form factor and also from a traditional method that analyzes the form factor using a fit ansatz. We compare our preliminary results obtained by these methods with previous lattice calculations and experiments.
}

\FullConference{The 41st International Symposium on Lattice Field Theory (LATTICE2024)\\
 28 July - 3 August 2024\\
Liverpool, UK\\}

\begin{document}
\maketitle

\section{Introduction}
The (mean-square) charge radius is a fundamental quantity that characterizes the structure of hadrons and describes the spread of the charge distribution. It is defined by the derivative of the electromagnetic form factor with respect to the momentum transfer. The form factor is obtained from the calculation of the matrix element of the electromagnetic current. In particular, the $\pi^{+}$ charge radius has recently been compared with experimental value with high precision. However, the results of the lattice calculations still have larger errors than the experimental value~\cite{FlavourLatticeAveragingGroupFLAG:2021npn}. There are four main systematic errors in the charge radius obtained from lattice calculations: chiral extrapolation, continuum extrapolation, finite volume effects, and fit ansatz.

\par

We have to choose a fit ansatz in the traditional method. For example, the charge radius can be obtained using the following procedure. First, the form factor of the $X$ meson, $F_X(Q^2)$, at a momentum transfer $Q^2$ is extracted by combining a proper three-point function and known factors. Second, we fit these data points by assuming a function for the form factor. Finally, using the fit result, the charge radius is obtained by calculating 
\begin{eqnarray}
    \expval{r_{X}^2}=-6\eval{\dv{F_{X}(Q^2)}{Q^2}}_{Q^2=0} .
\end{eqnarray}
In the second step of this procedure, the assumption for the fitting function and the selection of the fitting range cause the systematic error of the fit ansatz. 

\par

On the other hand, the model-independent method does not include the fit ansatz. The simplest method of the model-independent method is to calculate the difference, as in $(F_{X}(Q^2)-1)/Q^2$.
If the momentum transfer is small enough, we can obtain the first derivative of the form factor from it. The important point is that this method does not use any fit ansatz. Therefore it is called the model-independent method. However, this method can be affected by contamination from higher-order terms in the Taylor expansion of $F_X(Q^2)$ if the momentum transfer is large. In other words, the charge radius may be largely affected by it. Therefore, it is very important to reduce this contamination for precise calculations of the charge radius.

\par

We report our preliminary results for the $\pi^{+}$ and $K^{+}$ charge radii using a model-independent method~\cite{Feng:2019geu}, which was proposed to reduce the higher-order contamination, in the PACS10 configuration, whose spacetime volume is more than (10 fm)$^4$ at the physical point. We compare our results with those of previous lattice calculations and present our results are consistent with those.

\section{Overview of model-independent method}
In this section, we provide an overview of the model-independent method, explaining how a spatial moment is used to reduce the higher-order contamination. For a detailed discussion, see Refs.~\cite{Feng:2019geu, Sato:2022qee, Sato:2023vnb}.

\par

The simplest method of calculating the difference $(F_{X}(Q^2)-1)/Q^2$ might have a large contamination and is not suitable for precision calculations. To reduce the contamination, an important idea is calculating the spatial moment. For infinite volume limit, we derive
\begin{eqnarray}
    \eval{\dv{\tilde{C}_{\rm{3pt}}^{X}(t;p)}{p^2}}_{p^2=0} = \dfrac{1}{2}\sum_{x}x^2C_{\rm{3pt}}^{X}(t;x)
\end{eqnarray}
from the derivative of the Fourier transform of the three-point function of the $X$ meson, $C_{\rm{3pt}}^{X}(t;x)$. It means that the 1st-order momentum squared derivative equals the 2nd-order spatial moment. Therefore, we can suppress the contamination at large volumes by calculating the spatial moment of the three-point function. This method was used to calculate the slope of the Isgur-Wise function at zero recoil point~\cite{Aglietti:1994nx,Lellouch:1994zu} and the nucleon charge radius~\cite{Bouchard:2016gmc,Ishikawa:2021eut}. However, this method still has the contamination at a small volume.

\par

We can suppress higher-order contamination by cleverly combining the $x^2$ and $x^4$ moments even at a small volume, as in 
\begin{eqnarray}
    R_{\rm{MI}}^{X}(t) = \alpha_{1}{C_{\rm{3pt}}^{X}}^{(1)}(t) + \alpha_{2}{C_{\rm{3pt}}^{X}}^{(2)}(t) + h ,
    \label{eq:RatioMI}
\end{eqnarray}
where $\alpha_{1}, \alpha_{2}$ and $h$ are parameters set to cancel out the higher-order contamination and the $2n$-th moment of the three-point function,
\begin{equation}
    {C_{\rm{3pt}}^{X}}^{(n)}(t)
        =\frac{\displaystyle{\sum_{x}x^{2n}C_{\rm{3pt}}^{X}(t;x)}}{\displaystyle{\sum_{x}C_{\rm{3pt}}^{X}(t;x)}} . 
    \label{eq:moment_3ptfunc}
\end{equation}
This method was used to calculate the pion charge radius~\cite{Feng:2019geu}. In the following, we call this original model-independent method.

\par

We proposed a more improved version of the original model-independent method~\cite{Sato:2022qee, Sato:2023vnb}. In the reports, we presented that our improved method is effective at small lattice sizes. In this study, however, we use a large lattice size configuration called PACS10. Therefore, we do not have to use our improved method, while we use our method for a consistency check. Using these methods, we can determine the charge radius without fit ansatz.

\section{Lattice simulation and Preliminary results}
\subsection{Simulation parameters}
We apply these methods explained in the previous sections to the PACS10 configuration which is a 2+1 flavor gauge configuration generated by the PACS Collaboration~\cite{Ishikawa:2018jee}
with the nonperturbative $\order{a}$-improved Wilson action and the Iwasaki gauge action. This configuration has the characteristics of the physical point and a large volume exceeding $(\SI{10}{fm})^{4}$, so it is not necessary to consider chiral extrapolation or finite volume effects. And, since it is generated at 3 lattice spacings, we can perform a continuum extrapolation. In this study, we use the coarsest PACS10 configuration for the first step in this calculation. The ensemble parameters are shown in Table \ref{table:sim_param}.

\begin{table}[!h]
\begin{center}
\begin{tabular}{cccccccccc}\hline\hline
$\beta$ & $L^3\times T$ & $L$ [fm] & $a$ [fm] & $a^{-1}$ [GeV] & $m_\pi$ [MeV] & $m_K$ [MeV] & $N_{\rm conf}$ & $N_{\rm meas}$ \\
\hline
$1.82$ & $128^4$ & $10.9$ & $0.085$ & $2.316$ & $135$ & $497$ & 20 & 576  \\
\hline\hline
\end{tabular}
\end{center}
\caption{Details of the coarsest PACS10 configuration. The bare coupling ($\beta$), lattice size ($L^3\times T$), physical spatial extent ($L$[fm]), pion and kaon masses ($m_\pi, m_K$) are tabulated.
We represent $N_{\rm conf}$ and $N_{\rm meas}$ as the number of configurations
and the number of measurements per configuration, respectively.}
\label{table:sim_param}
\end{table}

\subsection{Correlation functions}
We use the one-dimensional two- and three-point functions defined by
\begin{eqnarray}
	C_{\rm{2pt}}^{X}(t;x_{1})&=&\sum_{x_{2},x_{3}}
	\matrixel{0}{\mathcal{O}_{X}(\vec{x},t){\mathcal{O}_{X}}^{\dag}(\vec{0},0)}{0} , 
    \label{eq:2ptfunc} \\
	C_{\rm{3pt}}^{X}(t;x_{1})&=&Z_{V}\sum_{\vec{y}}\sum_{x_{2},x_{3}}
	\matrixel{0}{\mathcal{O}_{X}(\vec{y},t_{\rm{sink}})V_{4}(\vec{x},t){\mathcal{O}_{X}}^{\dag}(\vec{0},0)}{0} ,
    \label{eq:3ptfunc}
\end{eqnarray}
where the interpolating operators $\mathcal{O}_{X}$, and the bare electromagnetic vector current $V_{4}$ are 
\begin{eqnarray}
    \mathcal{O}_{\pi^{+}}=\bar{d}\gamma_{5}u,\hspace{10pt}
    \mathcal{O}_{K^{+}}=\bar{s}\gamma_{5}u,\hspace{10pt}
    V_{4}=\sum_{f=u,d,s}Q_{f}\bar{\psi}_{f}\gamma_{4}\psi_{f} ,
\end{eqnarray}
and $Z_{V}$ is the renormalization factor of the vector current determined from that the value of the renormalized form factor in the zero-momentum transfer becomes the electric charge of the $X$ meson, as in $Z_{V}F_{X}^{\rm{bare}}(0)=+1$. The correlation functions are calculated with the $Z(2)\otimes Z(2)$ random source~\cite{Boyle:2008yd} and source-sink time separation $t_{\rm{sink}}=36$. The periodic boundary condition is imposed in the spatial directions of the correlation function, while the periodic and anti-periodic boundary conditions are employed in the temporal direction. The wrapping around effect~\cite{Aoki:2008ss,PACS:2017frz} is reduced by the average of the three-point function under these boundary conditions~\cite{PACS:2019hxd,Ishikawa:2022ulx}. Using $C_{\rm{3pt}}^{X}(t;x_{1})$ in Eq.~(\ref{eq:3ptfunc}), we calculate the moments of the three-point function as in Eq.~(\ref{eq:moment_3ptfunc}).

\par

The form factor for each momentum transfer squared is extracted from the time-independent part of the following ratio,
\begin{eqnarray}
    R_{\rm{FF}}^{X}(t;Q^2)=\qty(\dfrac{Z_{X}(0)}{Z_{X}(p)})\dfrac{2E_{X}(p)}{M_{X}+E_{X}(p)}\dfrac{\tilde{C}_{\rm{3pt}}^{X}(t;p)}{\tilde{C}_{\rm{3pt}}^{X}(t;\vec{0})}e^{(E_{X}(p)-M_{X})t} ,
    \label{eq:RatioFF}
\end{eqnarray}
where $p=\frac{2\pi}{L}n$ with $n$ and $L$ being an integer and the spatial extent, $E_{X}(p)=\sqrt{M_{X}^2+p^2}$, $Z_{X}(p)=\matrixel{0}{\mathcal{O}_{X}(\vec{0},0)}{E_{X}(p)}$, and the momentum transfer squared $Q^2=2M_{X}(E_{X}(p)-M_{X})$. $\tilde{C}_{\rm{3pt}}^{X}(t;p)$ is the momentum projection of the three-point function in Eq.(\ref{eq:3ptfunc}).

\subsection{Preliminary results}
\begin{figure}[htbp]
    \centering
    \begin{minipage}[b]{0.49\columnwidth}
        \centering
        \includegraphics[width=\linewidth]{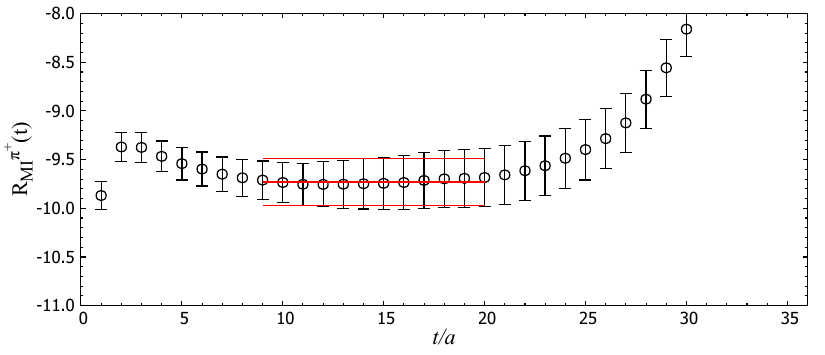}
    \end{minipage}
    \begin{minipage}[b]{0.49\columnwidth}
        \centering
        \includegraphics[width=\linewidth]{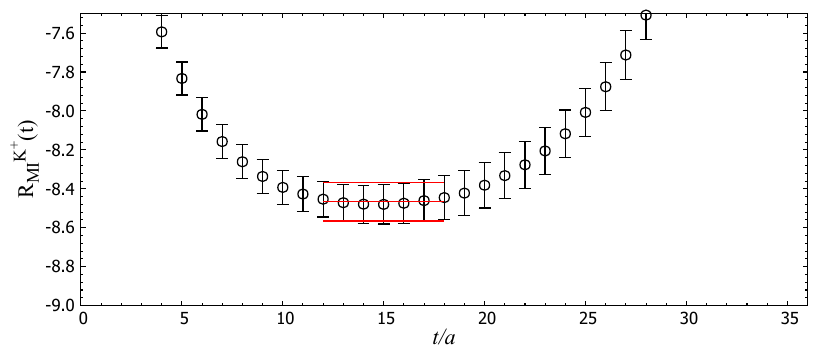}
    \end{minipage}
    \caption{Results of $R_{\rm{MI}}^{X}(t)$ in Eq.~(\ref{eq:RatioMI}) for $\pi^{+}$ meson (left) and $K^{+}$ meson (right). The solid red lines express the first derivative of the form factor obtained from a constant fit of a flat region.}
    \label{fig:model-independent_method_RMI}
\end{figure}

We obtain the charge radii for the $\pi^{+}$ and $K^{+}$ mesons using the traditional method with fit ansatz and the model-independent method. In the traditional method, the data of the form factor are obtained from the constant part of the ratio in Eq.~(\ref{eq:RatioFF}), where the amplitude $Z_{X}(p)$ and $E_X(p)$ are obtained from Eq.~(\ref{eq:2ptfunc}) after a proper momentum projection. The charge radii are calculated from the data points for the form factor using the fit ansatz of monopole, polynomial, z-expansion~\cite{Hill:2010yb}, and the NLO chiral perturbation theory (ChPT) formula\footnote{We use SU(2)-ChPT~\cite{Gasser:1983yg} for pion analysis and SU(3)-ChPT~\cite{Gasser:1984gg,Gasser:1984ux} for kaon.}. On the other hand, we calculate Eq.~(\ref{eq:RatioMI}) to obtain the charge radii directly without fit ansatz in the model-independent method, as shown in Fig.~\ref{fig:model-independent_method_RMI}. We see a time-independent region between the source and the sink time slices. The value of the first-order derivative of the form factor is extracted from this flat part.

\begin{figure}[htbp]
    \centering
    \begin{minipage}[b]{0.49\columnwidth}
        \centering
        \includegraphics[width=\linewidth]{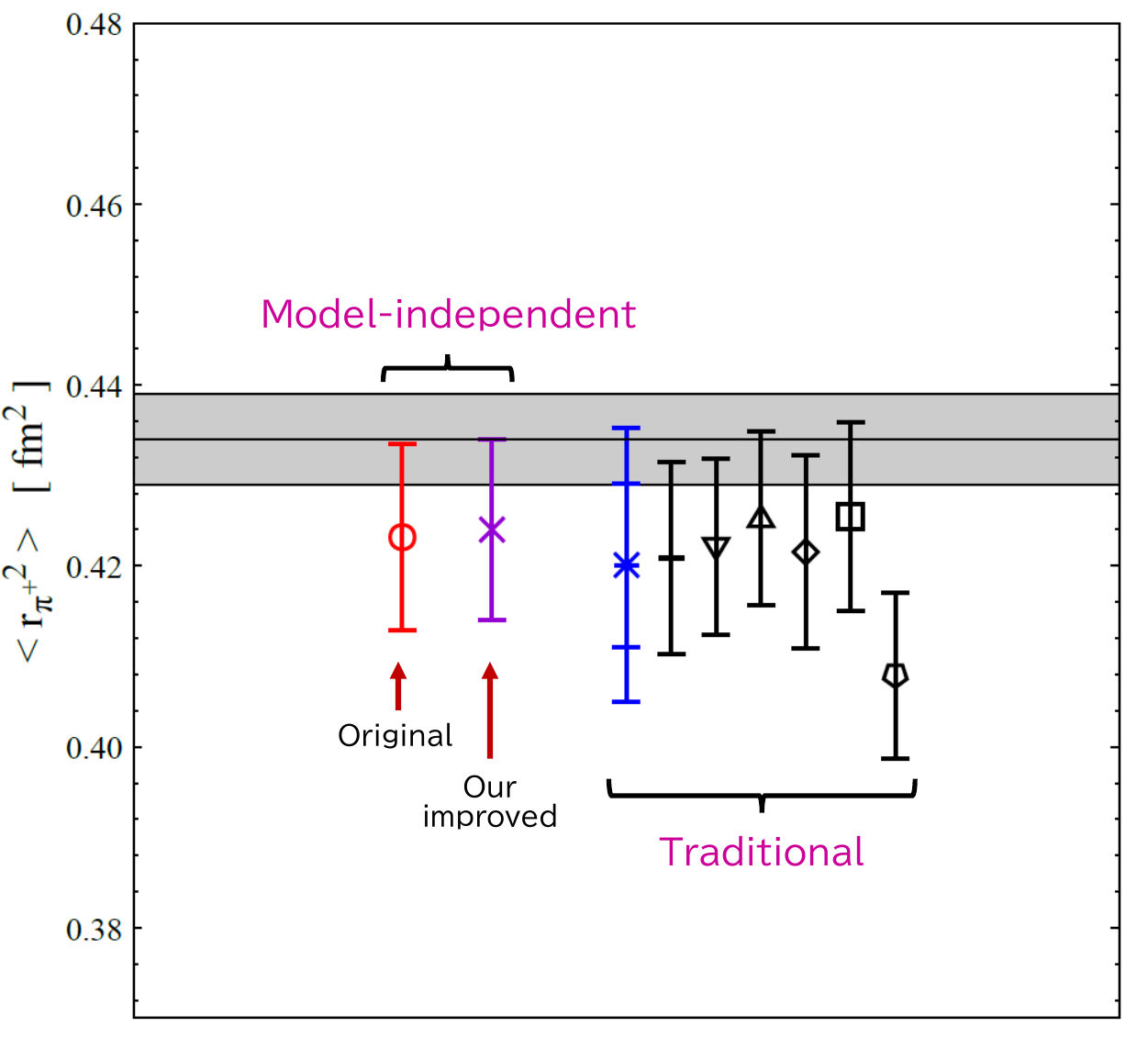}
    \end{minipage}
    \begin{minipage}[b]{0.49\columnwidth}
        \centering
        \includegraphics[width=\linewidth]{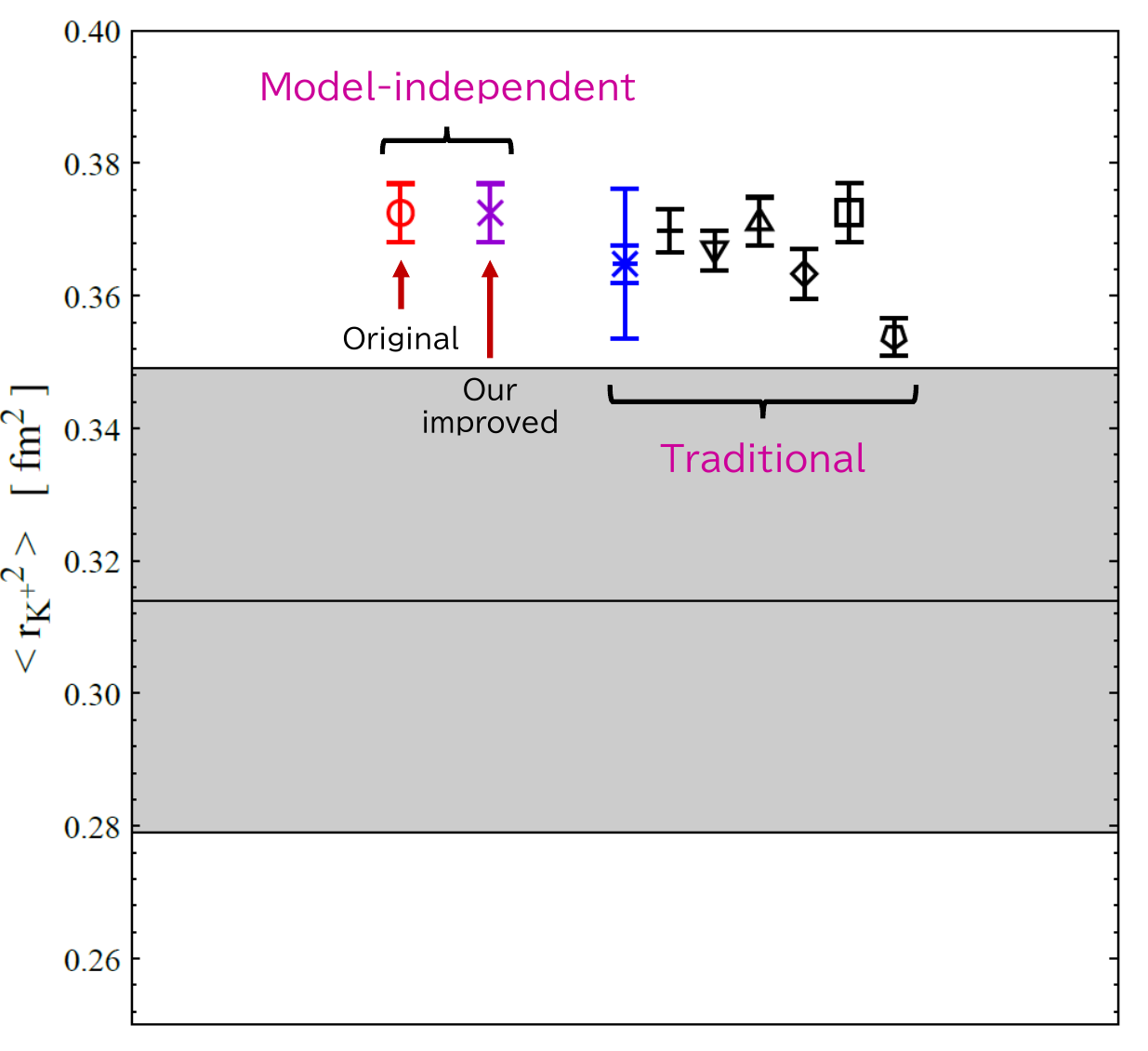}
    \end{minipage}
    \caption{Preliminary results for $\pi^{+}$ meson (left) and $K^{+}$ meson (right) charge radii obtained from traditional and model-independent methods. The black and blue symbols are the results of the traditional method. The black circle, inverted triangle, triangle, diamond, square, and crosse symbols are the results of using monopole, quadratic polynomial, cubic polynomial, quadratic z-expansion, cubic z-expansion, and ChPT NLO, respectively. The blue symbols are the result of considering systematic errors of fit ansatz. The red and purple symbols are the result of the original model-independent method and our improved model-independent method, respectively. The black line and gray band represent the central value and error of PDG22~\cite{ParticleDataGroup:2022pth}.}
    \label{fig:results_charge_radii}
\end{figure}

\par

The results obtained by each method are shown in Fig.~\ref{fig:results_charge_radii}. The black symbols in Fig.~\ref{fig:results_charge_radii} represent the results of the traditional method using the six functional forms for both $\pi^{+}$ and $K^{+}$. These data are reasonably consistent with each other within the error. However, there is variation in the central value depending on the fit ansatz, which is the systematic error of the fit ansatz. In this study, we evaluate the error of the traditional method as follows, as in Ref.~\cite{Sato:2023vnb}. The central value and statistical error are the weighted average of these results and the jackknife error of the central value, respectively. The systematic error is the maximum difference between the central value and the value on each form. We obtain the blue symbols in Fig.~\ref{fig:results_charge_radii} as the result of the traditional methods.

\par

We plot the results obtained from the original model-independent method and our improved method with log function~\cite{Sato:2023vnb} shown in the red and purple symbols in Fig.~\ref{fig:results_charge_radii}, respectively. They are consistent with those for the traditional methods. Furthermore, the model-independent method does not suffer from fit ansatz error and has smaller errors than the traditional method. In addition, our improved method agrees with the original model-independent method, because contamination is already well suppressed due to the large volume of the PACS10 configuration. This behavior is expected from our previous studies~\cite{Sato:2022qee, Sato:2023vnb}.

\par

All calculation results are consistent with the experimental values~\cite{ParticleDataGroup:2022pth} within the margin of error. In particular, we obtain $K^{+}$ charge radius with a smaller error than the experimental value. The error is about eight times smaller than the experimental value.

\par

\begin{figure}[htbp]
    \centering
    \begin{minipage}[b]{0.48\columnwidth}
        \centering
        \includegraphics[width=\linewidth]{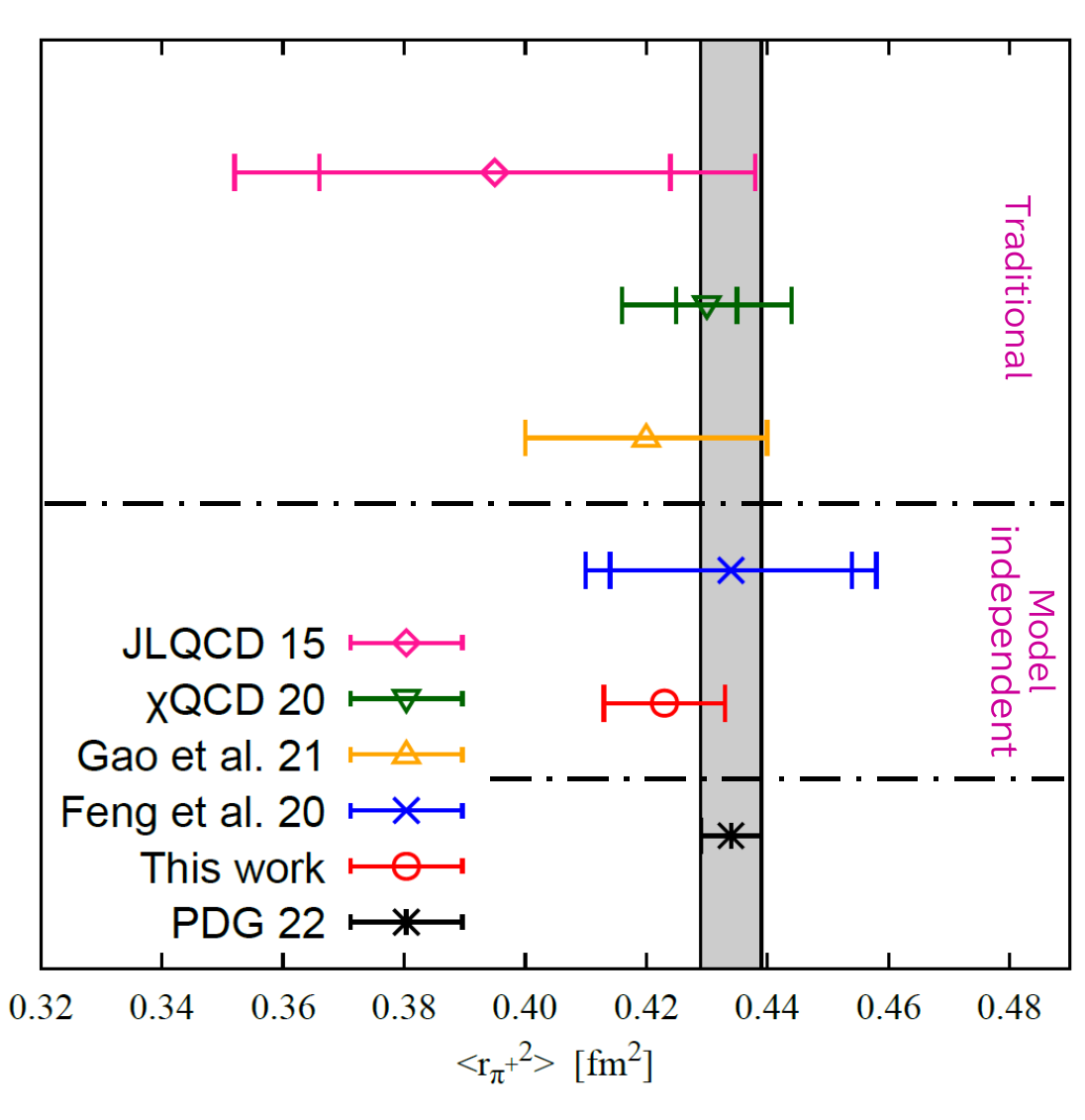}
    \end{minipage}
    \begin{minipage}[b]{0.49\columnwidth}
        \centering
        \includegraphics[width=\linewidth]{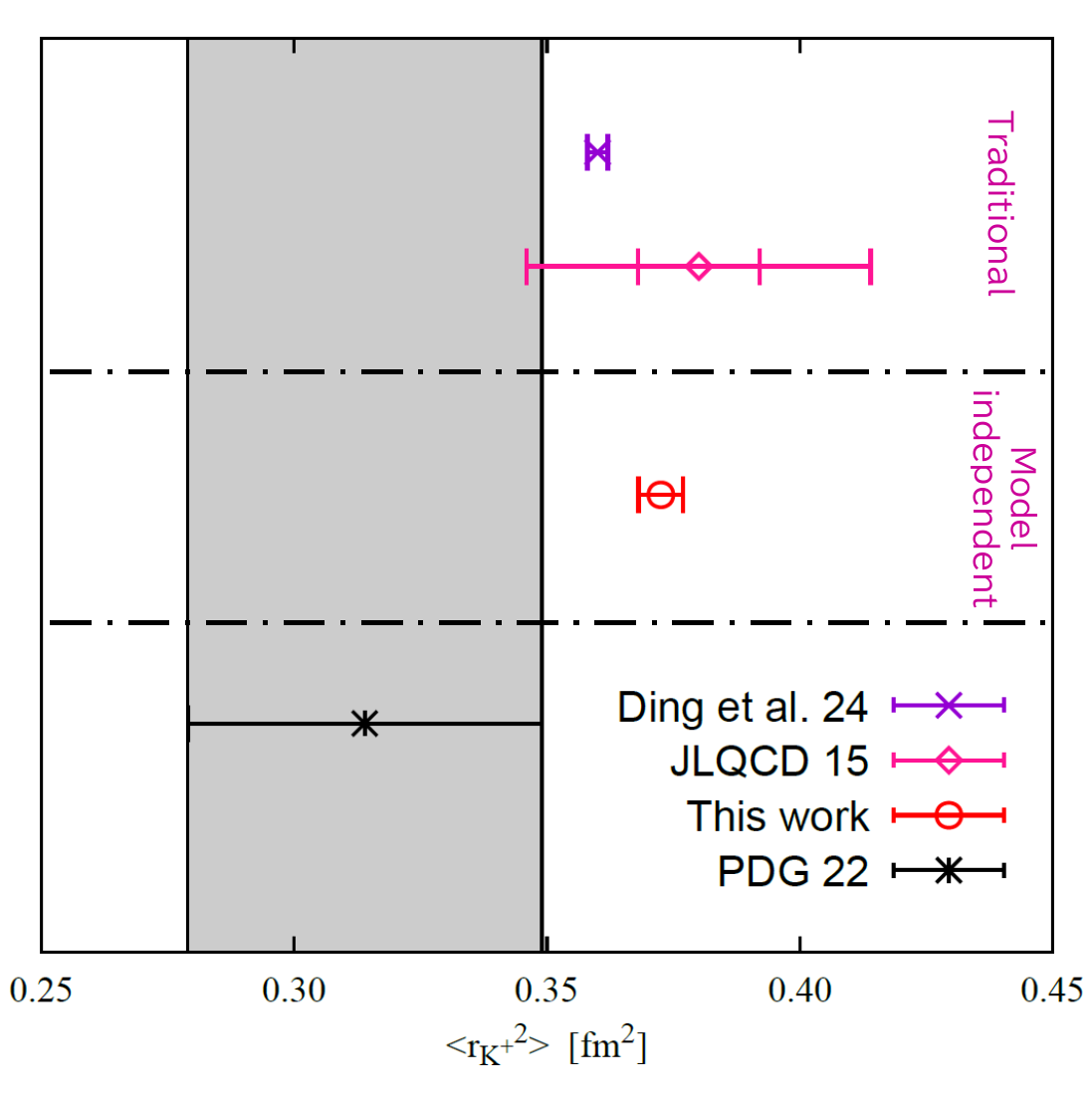}
    \end{minipage}
    \caption{The comparison of our preliminary result of the charge radius between those for previous lattice QCD calculations and PDG value. The left graph shows the $\pi^{+}$ charge radius and right graph is $K^{+}$ one. The colored symbols are the lattice calculations\cite{Feng:2019geu,Gao:2021xsm,Wang:2020nbf,Aoki:2015pba,Ding:2024lfj}, especially the red symbol is our preliminary result. The black line and gray band represent the central value and error of PDG22~\cite{ParticleDataGroup:2022pth}.}
    \label{fig:comparison_charge_radii}
\end{figure}
We compare our results, obtained using the model-independent method, with the results of previous lattice calculations and also the experimental value~(PDG22). From Fig.~\ref{fig:comparison_charge_radii}, we see that our results are reasonably consistent with these values.

\section{Summary}
We have presented the preliminary results of the charge radii of $\pi^{+}$ and $K^{+}$ mesons calculated with the coarsest PACS10 configuration at the lattice spacing of $\SI{0.085}{fm}$. This configuration is generated at the physical point on a $(\SI{10}{fm})^{4}$ volume.

\par

We have obtained the preliminary values of the charge radii for $\pi^{+}$ and $K^{+}$ obtained using a model-independent method:
\begin{eqnarray}
    \expval{r_{\pi^{+}}^2}=\SI{0.423(10)}{fm^2}, \hspace{15pt}
    \expval{r_{K^{+}}^2}=\SI{0.373(4)}{fm^2}    .
\end{eqnarray}
Our preliminary results are reasonably consistent with the previous lattice calculations and also the experimental values. In particular, our $K^{+}$ charge radius is more accurate than the experimental value. In this study, the systematic error of source-sink time separation and continuum extrapolation were not considered. Therefore, evaluating these errors are important tasks.

\section*{Acknowledgments}
Numerical calculations in this work were performed on Oakforest-PACS and Wisteria/BDEC-01 (Odyssey) in Joint Center for Advanced High Performance Computing, and on Cygnus in Center for Computational Sciences at the University of Tsukuba under Multidisciplinary Cooperative Research Program of Center for Computational Sciences, University of Tsukuba. The calculation employed OpenQCD system\footnote{http://luscher.web.cern.ch/luscher/openQCD/}. This work was supported in part by Grants-in-Aid for Scientific Research from the Ministry of Education, Culture, Sports, Science and Technology (No. 19H01892, 23H01195, 23K25891), MEXT as “Program for Promoting Researches on the Supercomputer Fugaku” (JPMXP1020230409), the JLDG constructed over the SINET5 of NII, JST SPRING, Japan Grant Number JPMJSP2124, and JST, the establishment of university fellowships towards the creation of science technology innovation, Grant Number JPMJFS2106.

\bibliographystyle{JHEP}
\bibliography{reference}

\end{document}